\documentclass[manuscript]{acmart}
\usepackage{bm}
\renewcommand\footnotetextcopyrightpermission[1]{} 
\begin{document}

\title{OFFER: A Motif Dimensional Framework for Network Representation Learning}


\author{Shuo Yu}
\affiliation{%
    \institution{Dalian University of Technology}
    \city{Dalian}
    \country{China}}
\email{y\_shuo@outlook.com}

\author{Feng Xia}
\affiliation{%
    \institution{Federation University Australia}
    \streetaddress{VIC 3353}
    \city{Ballarat}
    \country{Australia}}
\email{f.xia@ieee.org}

\author{Jin Xu}
\affiliation{%
    \institution{Dalian University of Technology}
    \city{Dalian}
    \country{China}}
\email{xujin0909@hotmail.com}

\author{Zhikui Chen}
\affiliation{
    \institution{Dalian University of Technology}
    \city{Dalian}
    \country{China}}
\email{zkchen@dlut.edu.cn}

\author{Ivan Lee}
\affiliation{%
    \institution{University of South Australia}
    \streetaddress{SA 5095}
    \country{Australia}}
\email{ivan.lee@unisa.edu.au}

%
%

\renewcommand{\shortauthors}{Shuo et al.}

\begin{abstract}
  Aiming at better representing multivariate relationships, this paper investigates a motif dimensional framework for higher-order graph learning. The graph learning effectiveness can be improved through OFFER. The proposed framework mainly aims at accelerating and improving higher-order graph learning results. We apply the acceleration procedure from the dimensional of network motifs. Specifically, the refined degree for nodes and edges are conducted in two stages: (1) employ motif degree of nodes to refine the adjacency matrix of the network; and (2) employ motif degree of edges to refine the transition probability matrix in the learning process. In order to assess the efficiency of the proposed framework, four popular network representation algorithms are modified and examined. By evaluating the performance of OFFER, both link prediction results and clustering results demonstrate that the graph representation learning algorithms enhanced with OFFER consistently outperform the original algorithms with higher efficiency.
\end{abstract}




\maketitle

\section{Introduction}


Network representation learning has been widely applied. The general idea of network representation learning is to convert nodes into vectors. By generating the structural information as well as node attributes together, network representation learning methods traverse the network structure to lower-dimensional vector space~\cite{mohan2019network}. The mechanism behind this is that directly connected node pairs are generally similar in the corresponding vector space, which leads to pairwise relationships that can be well represented in the corresponding low-dimensional vector space~\cite{wang2017malware}.


One of the most typical characteristics of network structure is that multivariate relationships can be profiled by higher-order structures~\cite{rossi2018interactive,Xu2020Muti}. However, in the corresponding low-dimensional vector space, such multivariate relationships are barely represented due to the limitation of vectors. Real-world networks consist of various complex relationships, in which most entities are in multivariate relationships with others~\cite{patra2018clustering,dey2019network}. Comparing with pairwise relationships, multivariate relationships are even more significant but complicated. For the previous methods, weights of edges are applied to evaluate the strength of pairwise relationships, but lack the ability to evaluate multivariate ones. 
Based on the network motif structures, multivariate relationships can be formulated and measured. 
Some previous studies realize that motif structure can help with network representation learning~\cite{rossi2018higher}, but the underlying reasons are not studied. Meanwhile, the feasibility of motifs is not discussed as well.

Driven by the significance of multivariate relationships, we propose OFFER (mOtiF dimensional Framework for nEtwork Representation learning) to enhance the abilities of network representation learning algorithms. This framework can be customized in choosing proper motifs when representing the network. We define two metrics based on the custom motifs: \textit{motif degree} and \textit{motif edge degree}. We refine the network adjacency matrix by motif degree of nodes, to help reduce the differences of node degrees that are introduced by higher-order structures. Then, we refine the learning process by employing the motif degree of edges, to strengthen the multivariate relationships in the network representation. The proposed method is examined in terms of link prediction and clustering. In summary, the contributions of this paper include:

\begin{itemize}
    
    \item \textbf{The design of motif dimensional framework for network representation:} We propose a motif dimensional framework to enhance the current network representation algorithms in formulating multivariate relationships. Enhanced by the proposed framework, the entities with multivariate relationships can be represented.

    \item \textbf{The discovery of the correlation between motif structures and multivariate relationships:} The mechanism of OFFER makes it possible to reinforce strong relationships as well as multivariate relationships, thus it is universally applicable for such kind of network representation methods.
    
    \item \textbf{The efficiency of OFFER:} The experimental results show that the enhanced methods generally lead to superior network representation performance. All enhanced methods outperform their original network representation algorithms with higher accuracy in both prediction and clustering capabilities.
    
\end{itemize}

\section{MODEL DESIGN}
\label{sec:md}
To characterize the frequent multivariate relationship structure in the network, we use the most common three-order triangle motif.

\begin{definition}
    \textbf{Network Motif.} For the graph $G = (V, E)$, the network motif $M$ is a special subgraph structure in $G$ that satisfies three characteristics: low-order nature, high-frequency nature, and real mapping nature.
\end{definition}

\begin{figure}[!b]
    \centerline{\includegraphics[width=0.38\textwidth]{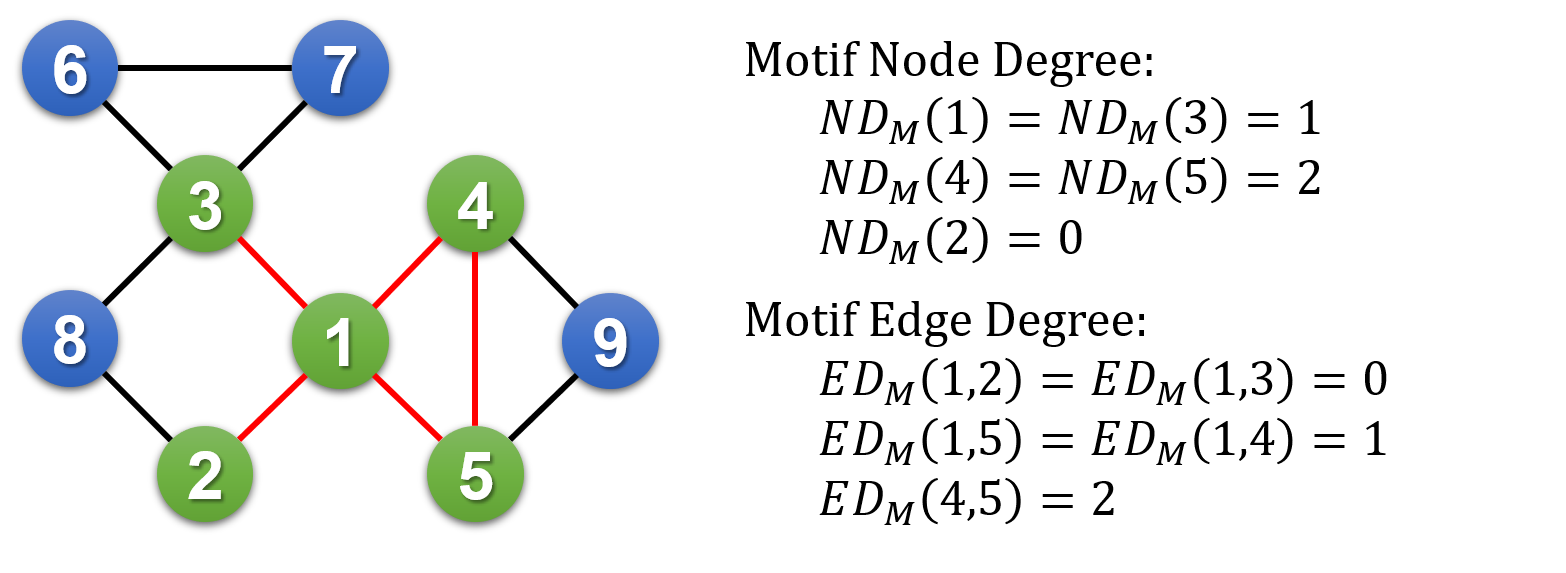}}
    \caption{An example to explain triangle MND and MED.}
    \label{fig:mexp}
\end{figure}

\begin{figure}[!t]
    \centering
    \includegraphics[width=0.47\textwidth]{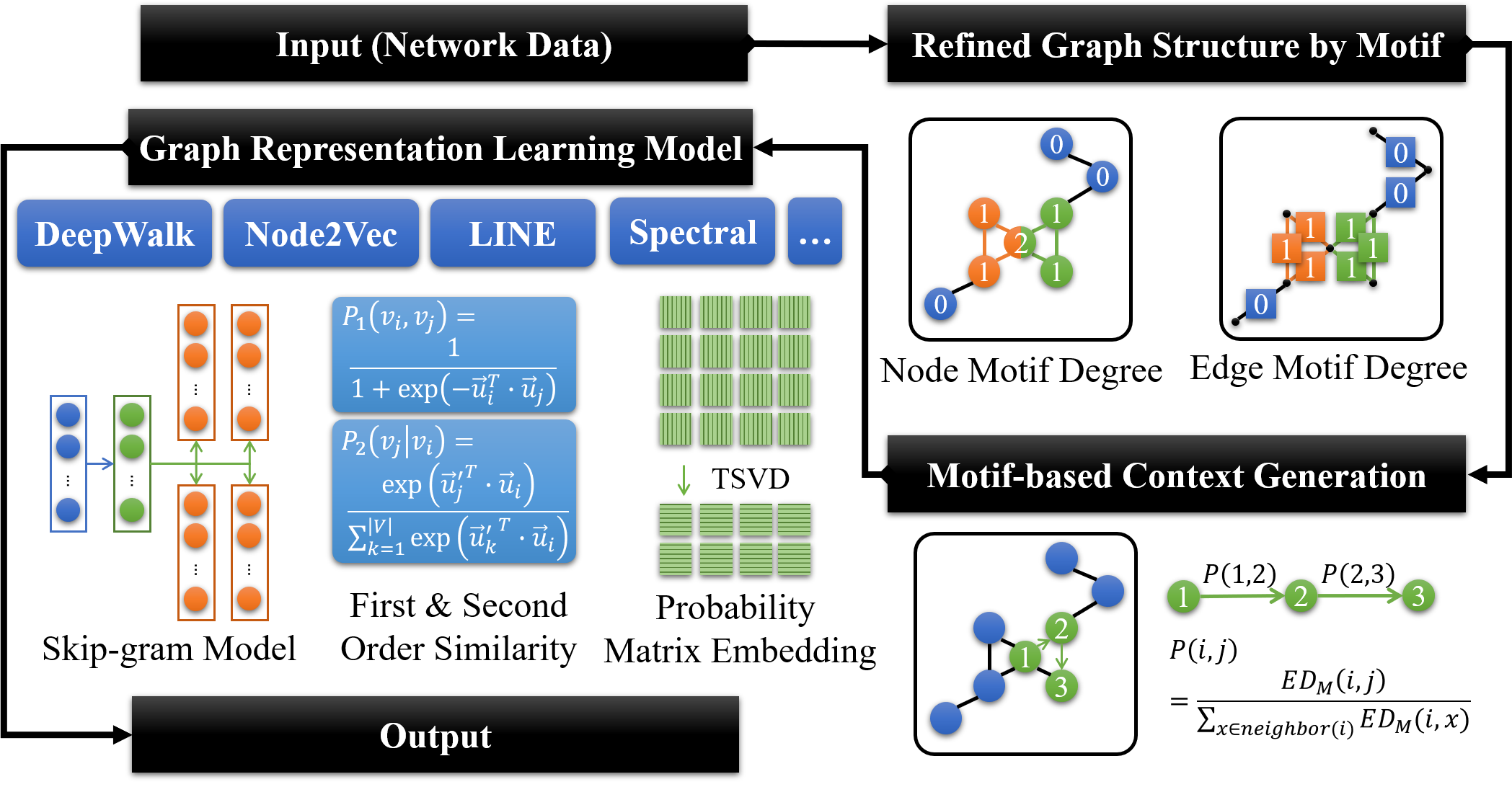}\\
    \caption{The framework of OFFER.}
    \label{fig:framework}
\end{figure}

Specifically, the low-order nature means that the order of $M$ is generally not high (not over 8). The high-frequency nature means that the frequency of $M$ in the real network is much higher than that of the corresponding random network. The real mapping nature means that $M$ can always find the actual meaning in the real network. Because of these natures of $M$, especially the real mapping nature, motifs can be used to characterize specific multivariate relationship structures in the real network. Among various network motifs, triangle motifs are more popular, because they are widely present in various real networks and have a strong reality mapping.

To characterize motif features on network nodes or edges, we define the motif degree of each node and edge.
\begin{definition}
    \textbf{Motif Node Degree (MND).} For a given motif $M$, $ND_M(i)$ represents the motif degree of node $i$, and $ND_M(i)=n$ means node $i$ is involved in a number of $n$ motifs.
\end{definition}
\begin{definition}
    \textbf{Motif Edge Degree (MED).} For a given motif $M$, $ED_M(i,j)$ represents the motif degree of edge $(i,j)$, and $ED_M(i,j)=n$ means edge $(i,j)$ is involved in a number of $n$ motifs.
  \end{definition}
  
Figure~\ref{fig:mexp} illustrates these two concepts. The formula on the right represents the triangle motif degree information of the corresponding nodes (green) and edges (red) on the left. Comparing with traditional node degree, MND and MED better reflect various network attributes such as significant network connectivity, robustness, and centrality. We refer to the analysis method using motif degree information instead of traditional degree information as the ``motif-dimensional analysis method''.

We define a motif-biased adjacency matrix $A_M$ for $G$ to refine the total network. This matrix is constructed based on MND. To ensure network connectivity, we define $A_M$ as shown in Equation~(\ref{eq:am}).
\begin{small}
  \begin{equation}
    A_M(i,j)=
    \begin{cases}
      1, & \text{if $(i,j) \in E$, $ED_M(i,j)=0$}\\
      0, & \text{if $(i,j) \notin E$}\\
      1+\frac{ED_M(i,j)}{|V_{M}|}, &\text{if $(i,j) \in E$,    $ED_M(i,j)\neq0$}
    \end{cases}
    \label{eq:am}
  \end{equation}
\end{small}
Wherein, $|V_{M}|$ represents the number of nodes in $M$.

The general framework is shown in Figure~\ref{fig:framework}. For a real network, OFFER first calculates its MND and MED matrix. Then, OFFER converts them into an appropriate network representation matrix, such as $A_M$. Finally, OFFER deploys the improved network representation learning algorithm.
As an example, for the network representation learning algorithm based on the transition probability matrix $P$, OFFER refine the learning process in network representation by using MED. Consequently, this will lead to a fact that edges with larger $ED_M(i,j)$ will achieve more learning weight. That is, if we denote the neighbor nodes of $i$ as id $neighbor(i)$, then the probability from node $i$ to node $j$ is shown in Equation~(\ref{eq:ptrans}).

\begin{equation}
P_{trans}(i,j)=\frac{ED_M(i,j)}{\sum_{x\in{neighbor(i)}}ED_M(i,x)}
\label{eq:ptrans}
\end{equation}

\section{EXPERIMENTS}
\label{sec:exp}

\begin{figure*}[!t]
    \begin{minipage}{0.32\textwidth}
        \centerline{\includegraphics[width=5.3cm]{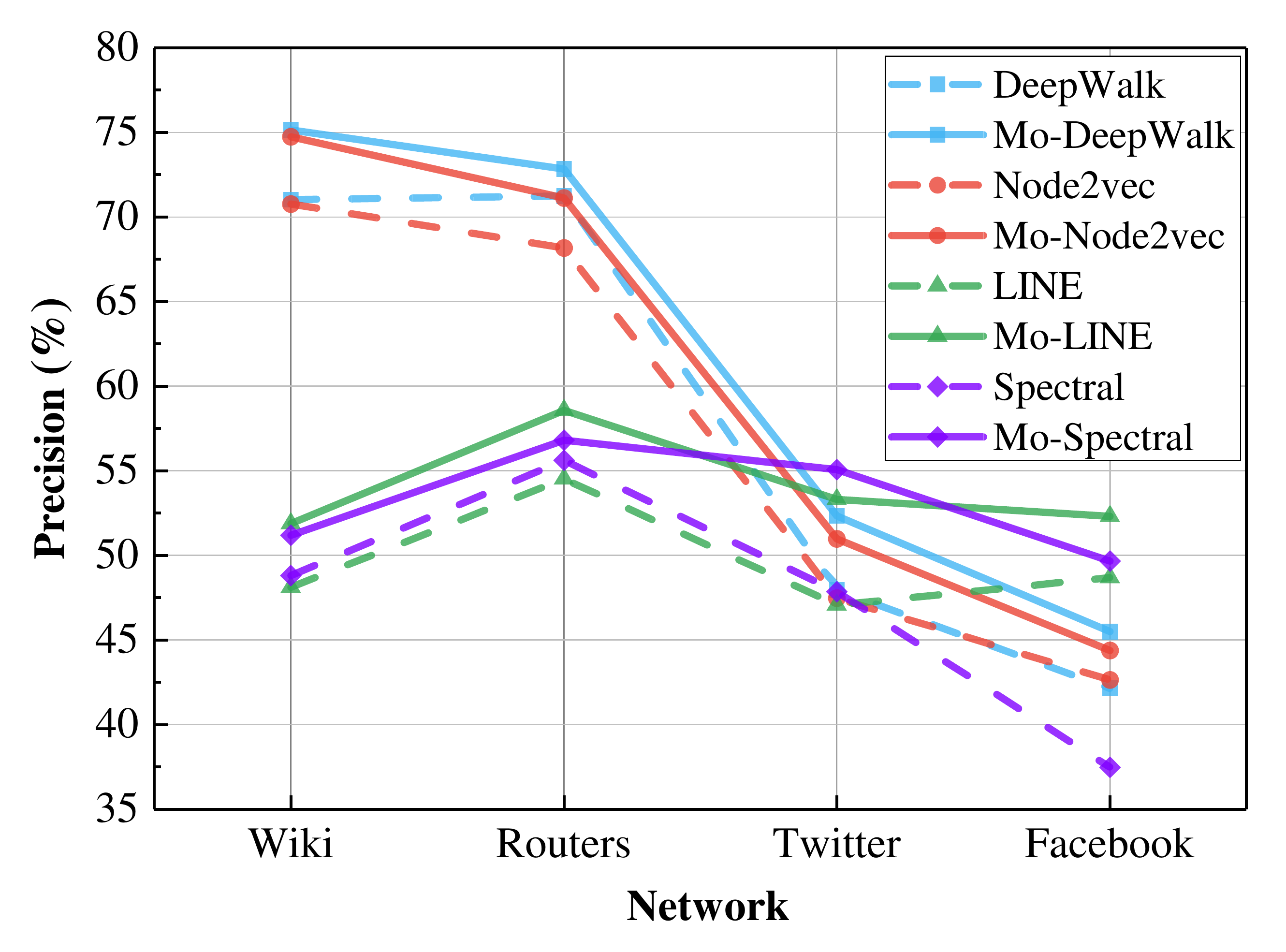}}
        \label{fig:Precision}
        \centering{(a) Precision}
    \end{minipage}
    \begin{minipage}{0.32\textwidth}
        \centerline{\includegraphics[width=5.3cm]{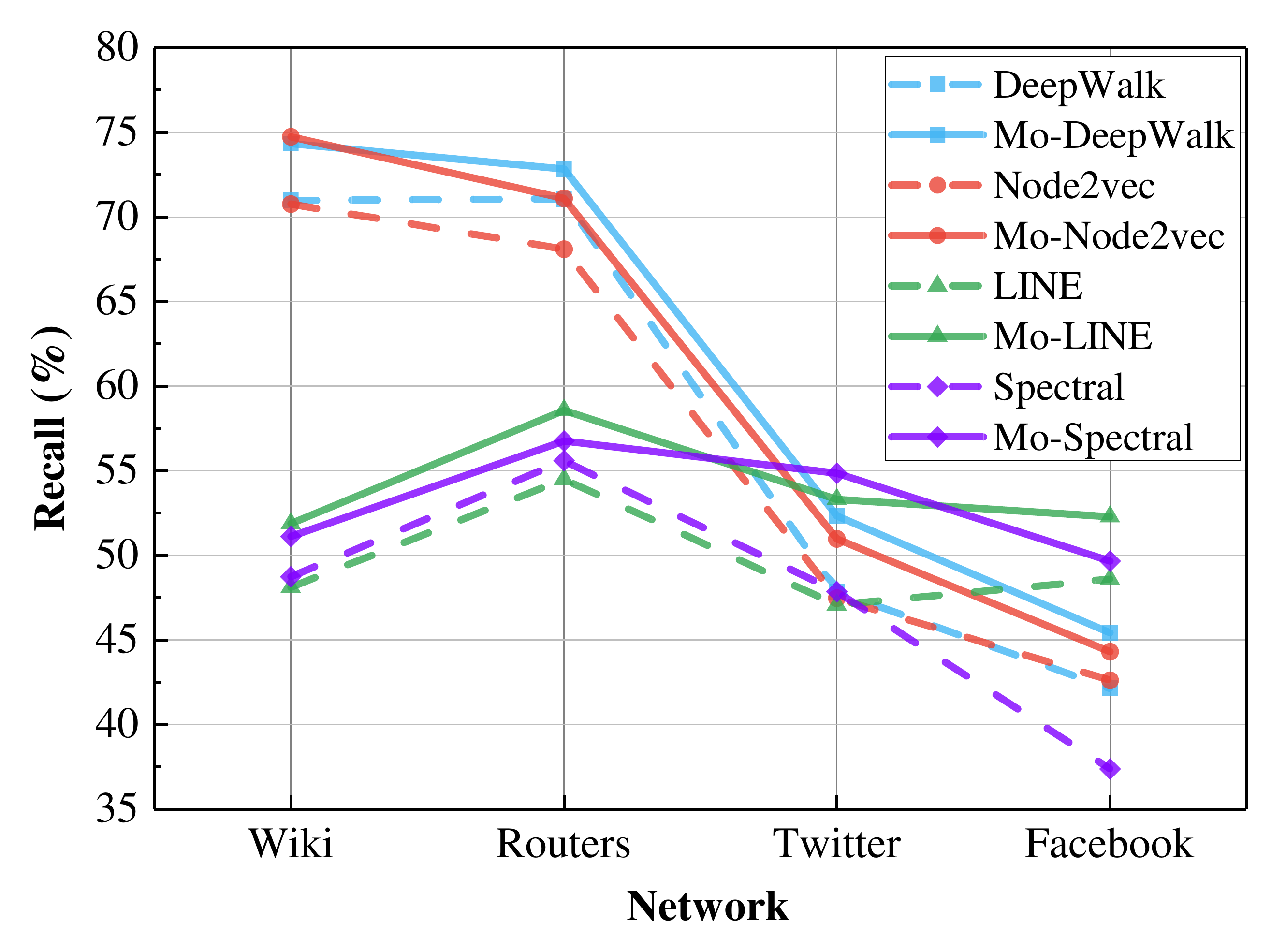}}
        \label{fig:Recall}
        \centering{(b) Recall}
    \end{minipage}
    \begin{minipage}{0.32\textwidth}
        \centerline{\includegraphics[width=5.3cm]{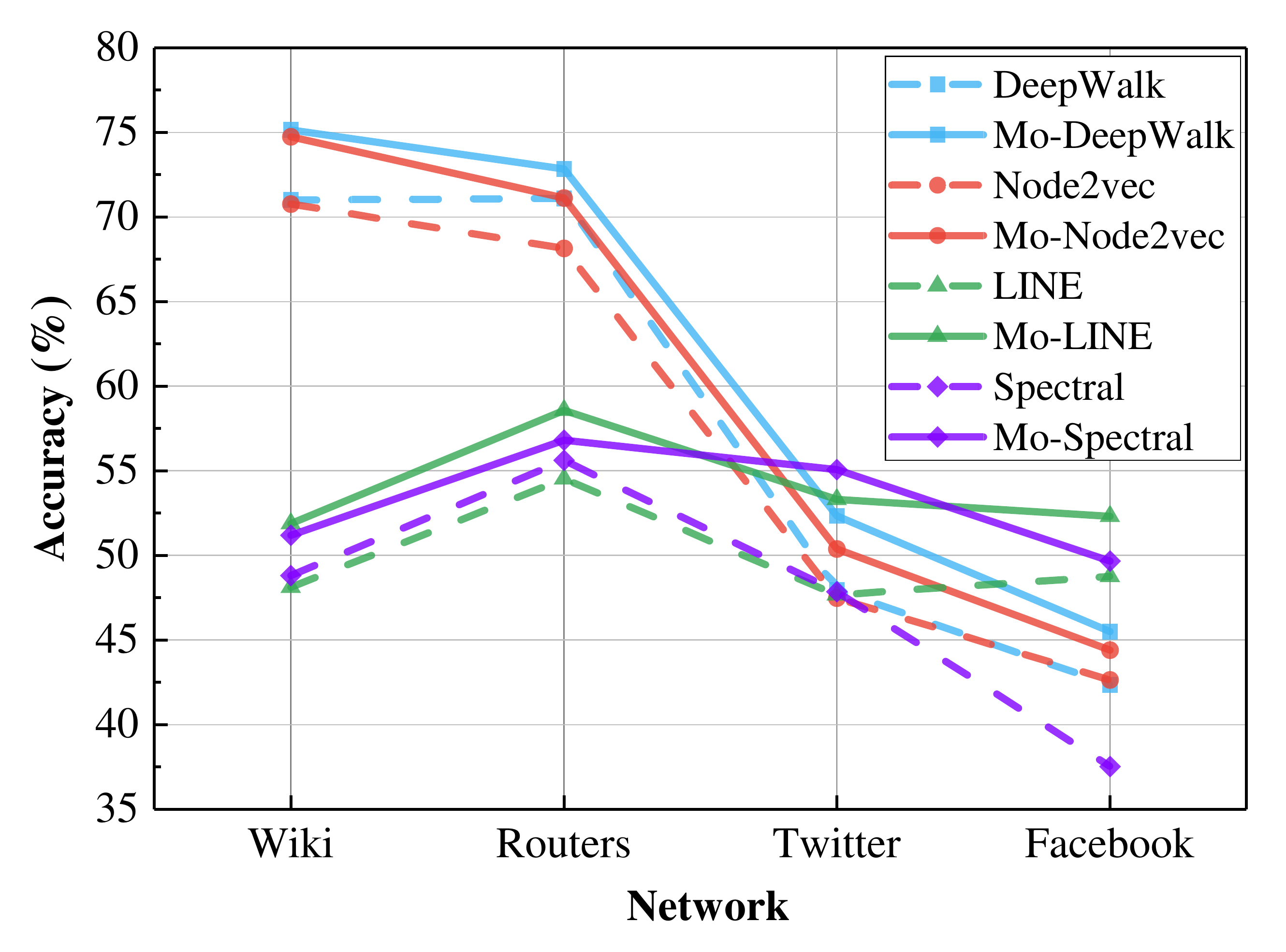}}
        \label{fig:Accuracy}
        \centering{(c) Accuracy}
    \end{minipage}
    \vfill
    \begin{minipage}{0.32\textwidth}
        \centerline{\includegraphics[width=5.3cm]{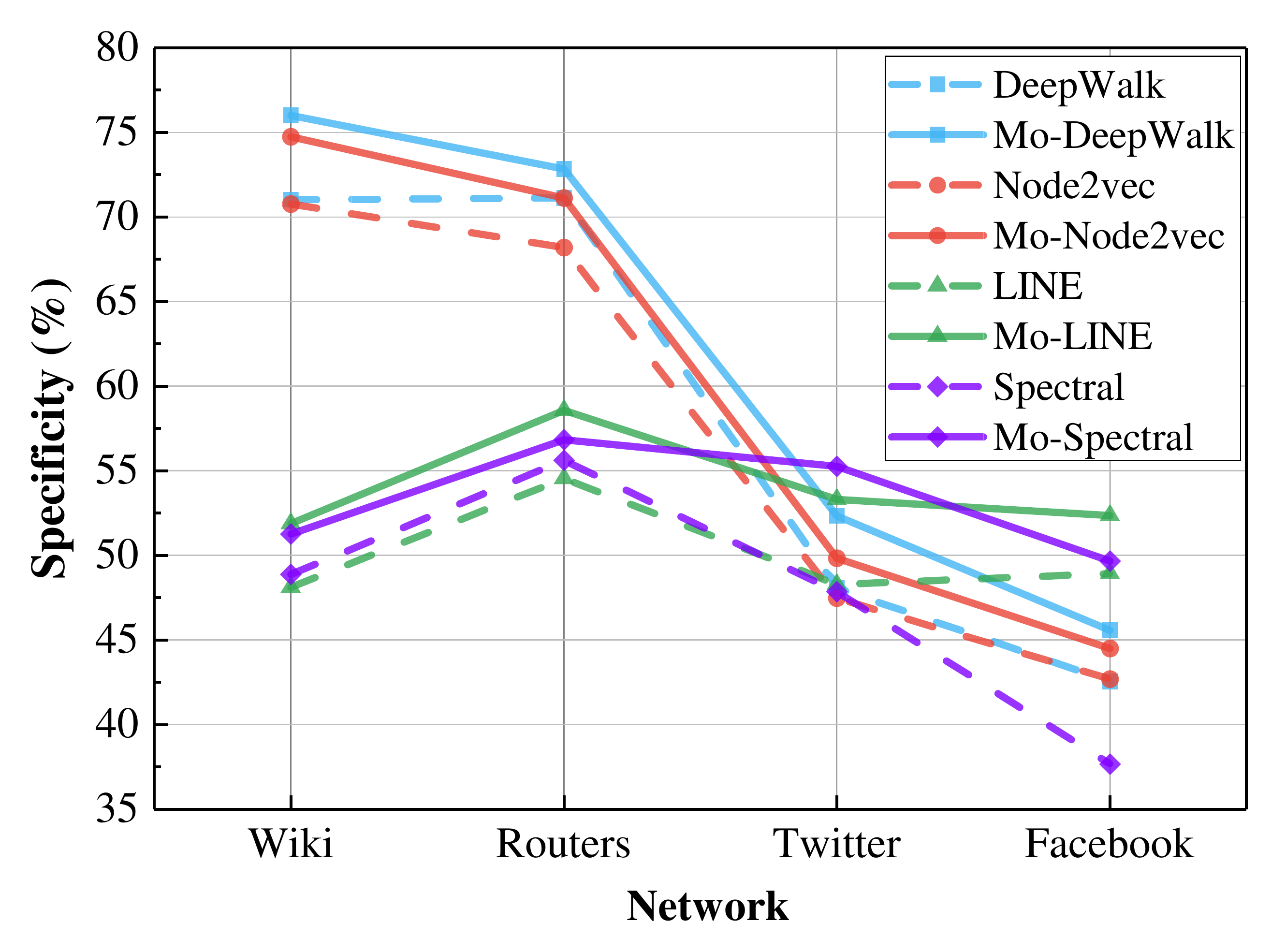}}
        \label{fig:Specificity}
        \centering{(d) Specificity}
    \end{minipage}
    \begin{minipage}{0.32\textwidth}
        \centerline{\includegraphics[width=5.3cm]{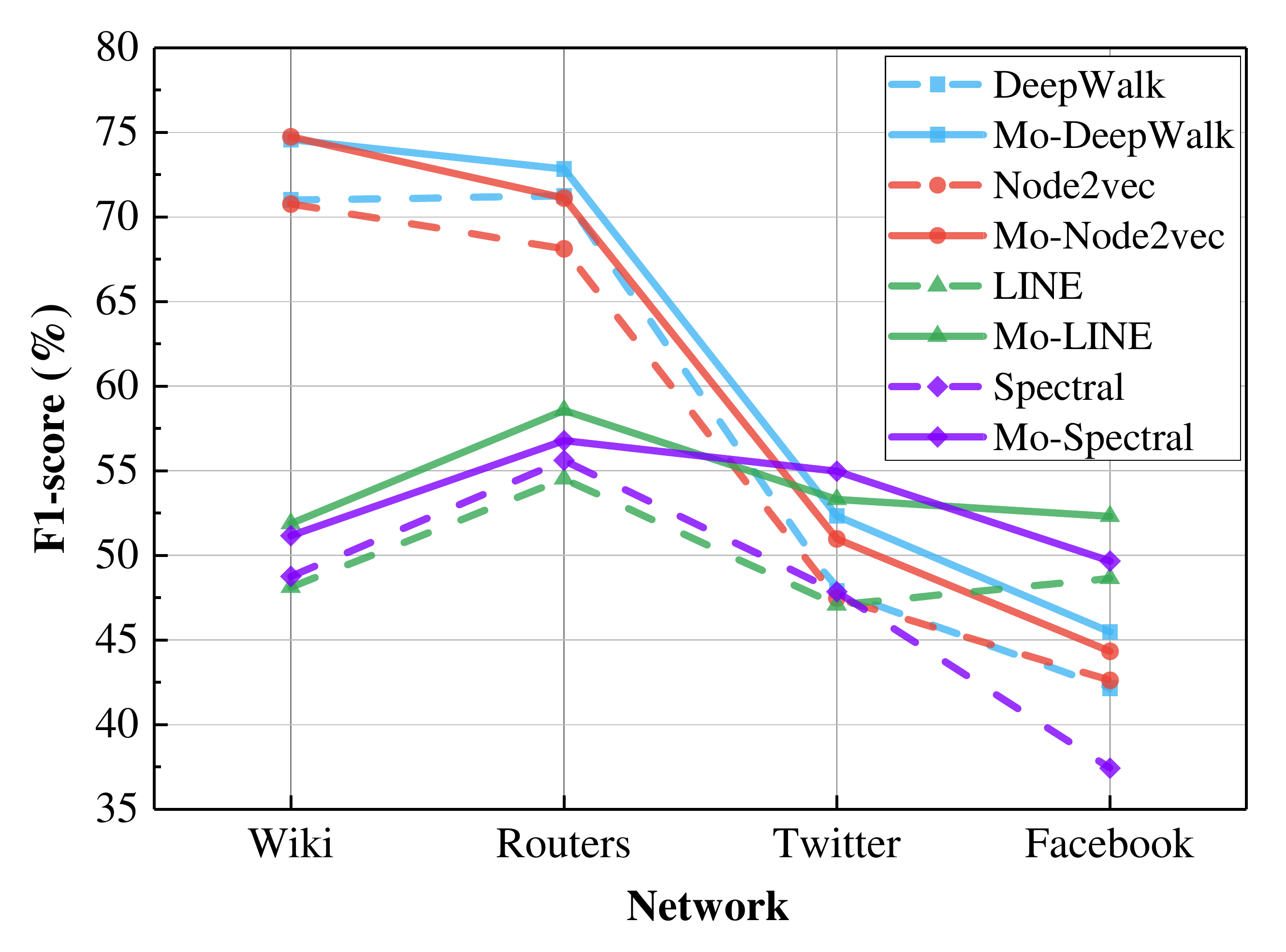}}
        \label{fig:F1-score}
        \centering{(e) F1-score}
    \end{minipage}
    \begin{minipage}{0.32\textwidth}
        \centerline{\includegraphics[width=5.3cm]{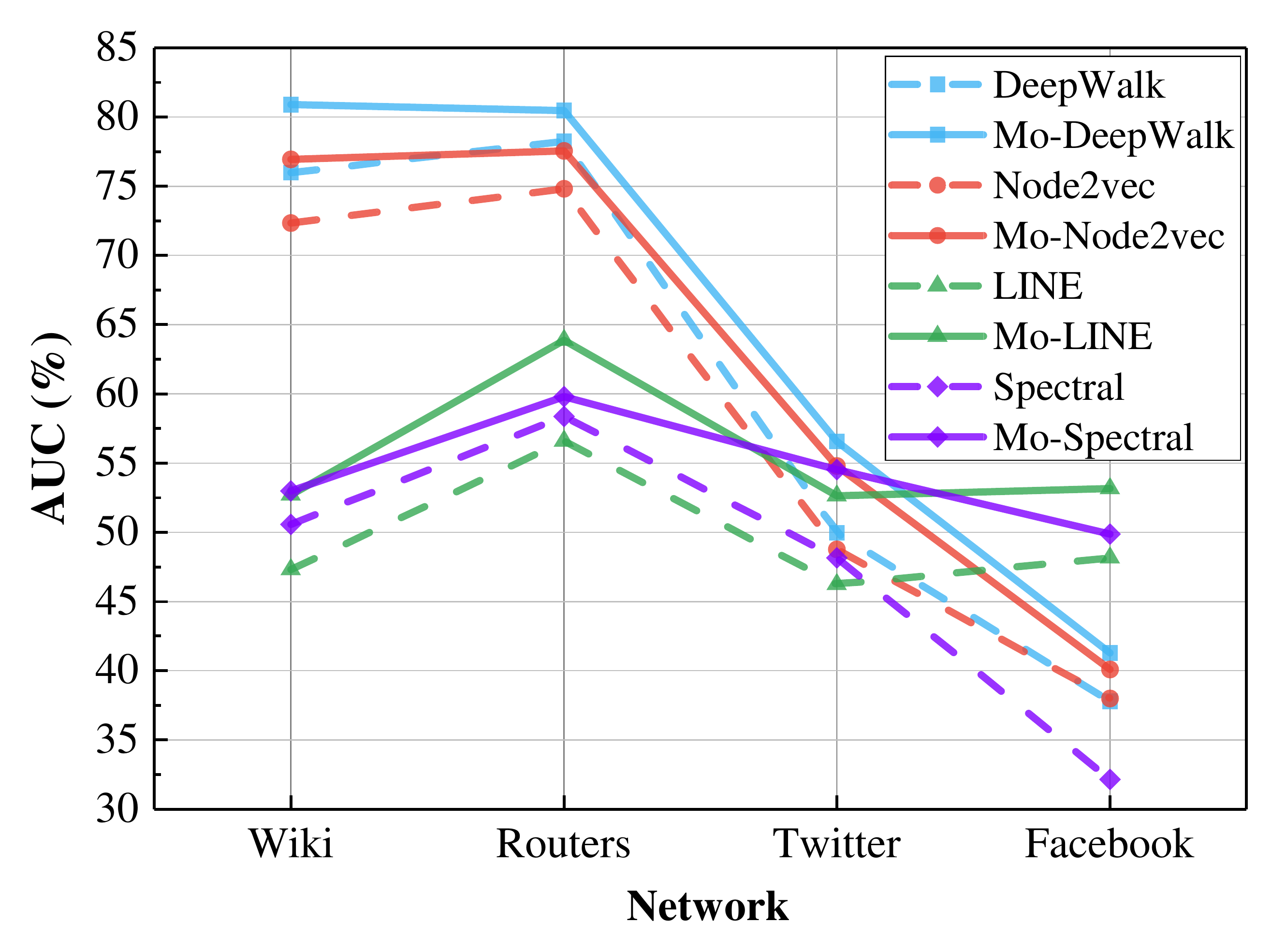}}
        \label{fig:AUC}
        \centering{(f) AUC}
    \end{minipage}
    
    \caption{Six indicators comparison in four networks.}
    \label{fig:PAR}
\end{figure*}

\subsection{Data Sets}

Four network datasets are selected for performance evaluation of link prediction, and these datasets include \textbf{Wiki}\footnote{\url{http://nrvis.com/download/data/soc/soc-wiki-Vote.zip}}, \textbf{Routers}\footnote{\url{http://nrvis.com/download/data/tech/tech-routers-rf.zip}}, \textbf{Twitter}\footnote{\url{http://networkrepository.com/rt-twitter-copen.php}}, and \textbf{Facebook}\footnote{\url{http://networkrepository.com/ego-facebook.php}}. The records in these datasets can be used to generate undirected and unweighted networks. For the clustering scenario, we use \textbf{Twitter} social dataset and \textbf{Routers} dataset. Meanwhile, we also use two new datasets, namely, \textbf{Hamsterster}\footnote{\url{http://networkrepository.com/soc-hamsterster-vote.php}} and \textbf{Openflights}\footnote{\url{http://networkrepository.com/inf-openflights.php}}. Both have higher network density to show strong universality of our method in various network environments. Basic properties of these networks are shown in Table~\ref{tab:info}. Wherein, $\bm{|V|}$ is the number of nodes, $\bm{|E|}$ is the number of edges, $\bm{\max{(d)}}$ is the max node degree, $\bm{\bar{d}}$ is the average node degree and $\bm{\rho}$ is the density.

\begin{table}[!b]
    \centering
    \caption{The statistic information of experimental datasets}
    \resizebox{0.43\textwidth}{!}{
        \begin{tabular}{c|ccccc}
            \toprule
            \textbf{Network}   & $\bm{|V|}$ & $\bm{|E|}$ & $\bm{\max{(d)}}$ & $\bm{\bar{d}}$ & $\bm{\rho}$ \\
            \midrule
            \textbf{Wiki}        & 889           & 2914          & 102           & 6.5556        & 0.007383    \\
            \textbf{Routers}     & 2113          & 6632          & 109           & 6.2773        & 0.002972    \\
            \textbf{Twitter}     & 761           & 1029          & 37            & 2.7043        & 0.003558     \\
            \textbf{Facebook}    & 2888          & 2981          & 769           & 2.0644        & 0.000715   \\
            \textbf{Hamsterster} & 2426          & 16630         & 273           & 13.7098       & 0.005654    \\
            \textbf{Openflights} & 2939          & 15677         & 242           & 10.6682       & 0.003631   \\
            \bottomrule
    \end{tabular}}
    \label{tab:info}%
\end{table}

\begin{figure}[!b]
    \centering
    \includegraphics[width=0.38\textwidth]{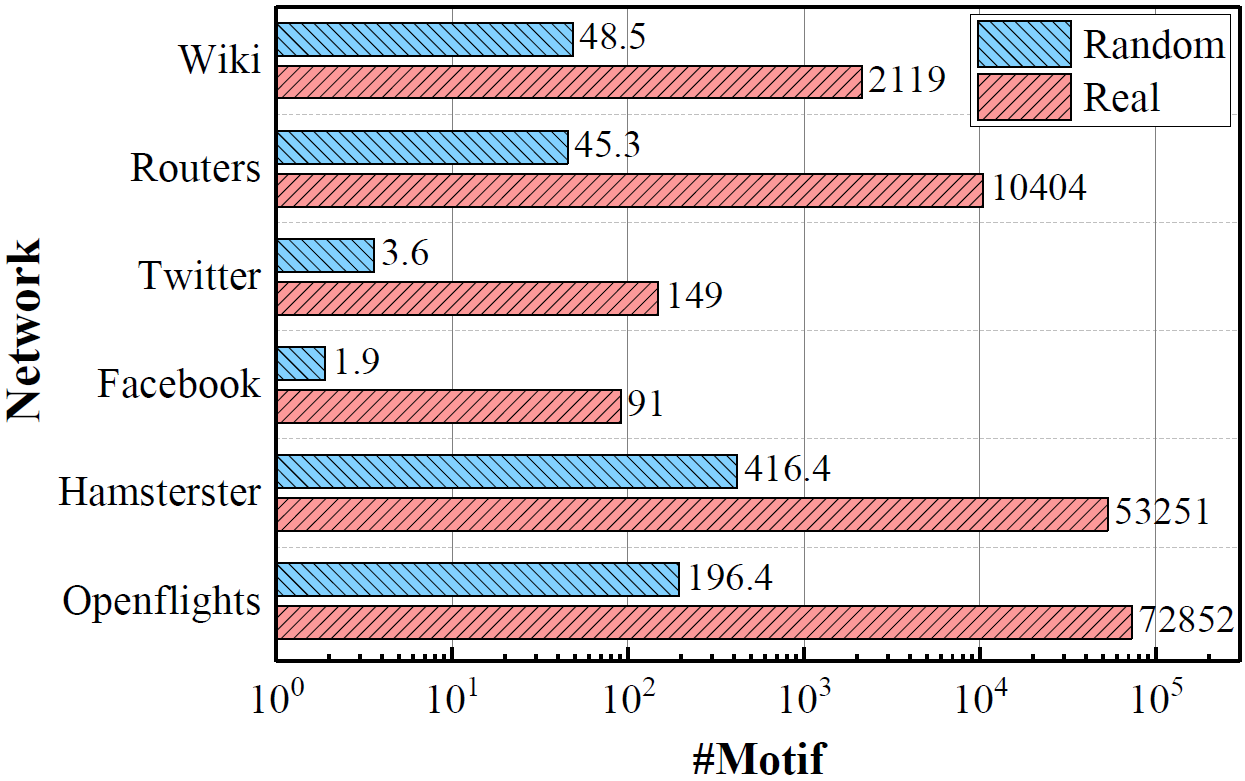}\\
    \caption{The triangle motif numbers in 6 datasets.}
    \label{fig:TC}
\end{figure}

In these network environments, triangle motifs have different realistic mapping meanings, which also represent different multivariate relationships. In social networks (Facebook, Twitter, Hamsterster), they are employed to represent ternary closures. In Openflights, they correspond to connecting flights. As for Routers, they represent the routing structures of network devices. Due to the realistic significance, we specifically make statistics of triangle motif distributions among the above-mentioned datasets and their corresponding random networks (shown in Figure~\ref{fig:TC}).

\subsection{Evaluation Metrics}

\textbf{Link Prediction. }
We use cosine similarity in the link prediction based on previous work~\cite{de2017community}, wherein the cosine similarity of two vectors $X,Y$ is calculated by the following formula:
$CosSim(X,Y) = \vec{x} \cdot \vec{y} / \|x\| \cdot \|y\|$.

Six metrics (AUC, Accuracy, Precision, Recall, Specificity, and F1-score) are employed to verify the performance of link prediction:
\textbf{AUC} reflects the overall performance of the proposed model, which refers to the area under ROC (Receiver Operating Characteristic) curve. \textbf{Accuracy} reflects the right prediction results ratio of the total samples, which is calculated by $ Accuracy = (TP+TN) / (TP+TN+FP+FN) $. \textbf{Precision} reflects how many of the prediction positive results are positive samples, defined as $ Precision = TP / (TP+FP) $. \textbf{Recall} evaluates the recognition ratio of the positive sample, defined as $ Recall = TP / (TP+FN) $. \textbf{Specificity} evaluates the recognition ratio of the negative sample, defined as $ Specificity = TN / (TN+FP) $. \textbf{F1-score} is calculated based on Precision and Recall, defined as:
$F1$-$score=(2 \times Precision \times Recall) / (Precision + Recall)$.

\begin{figure}[!t]
    \begin{minipage}{0.23\textwidth}
        \centerline{\includegraphics[width=3.5cm]{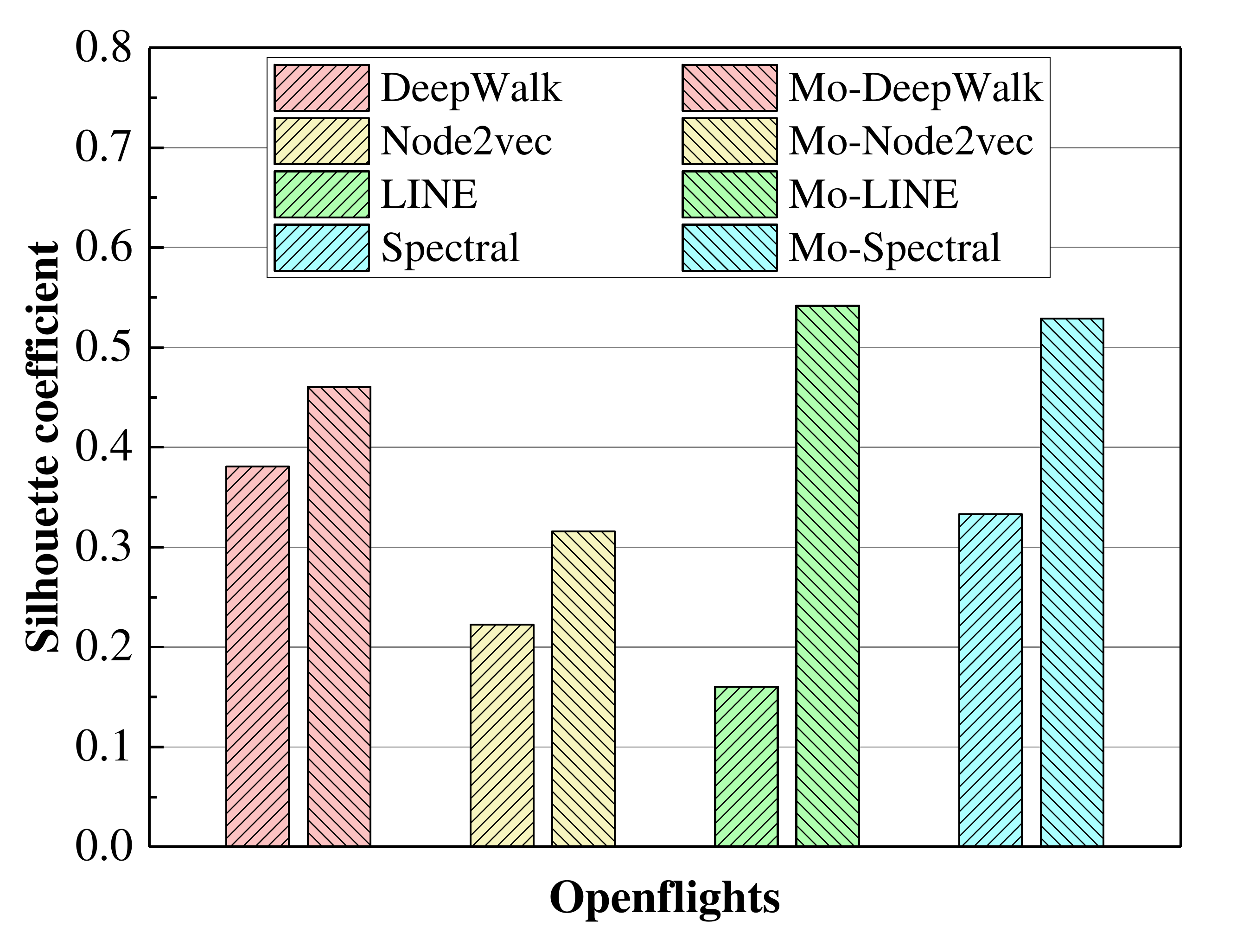}\label{fig:scopen}}
    \end{minipage}
    \begin{minipage}{0.23\textwidth}
        \centerline{\includegraphics[width=3.5cm]{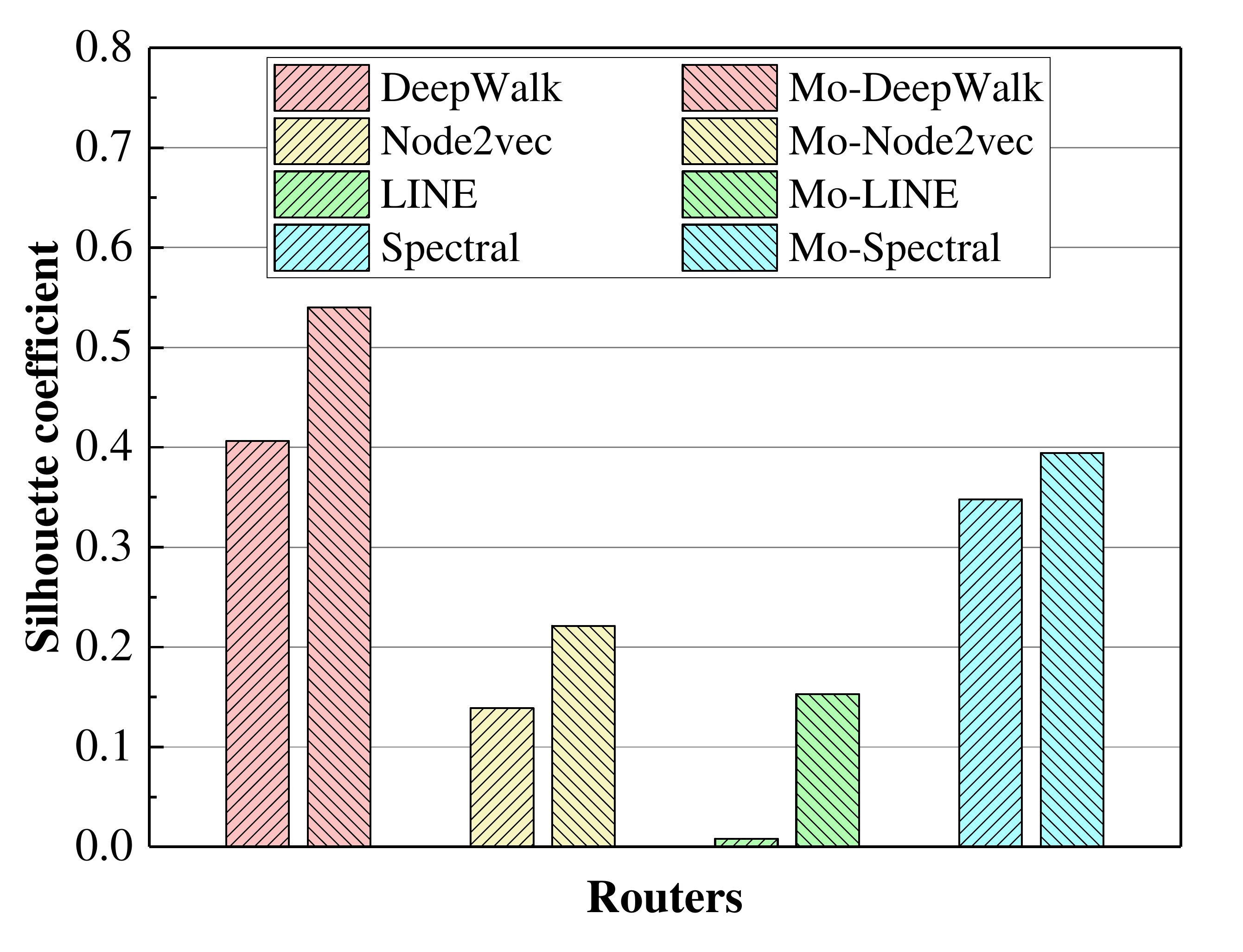}\label{fig:scrouters}}
    \end{minipage}
    \vfill
    \begin{minipage}{0.23\textwidth}
        \centerline{\includegraphics[width=3.5cm]{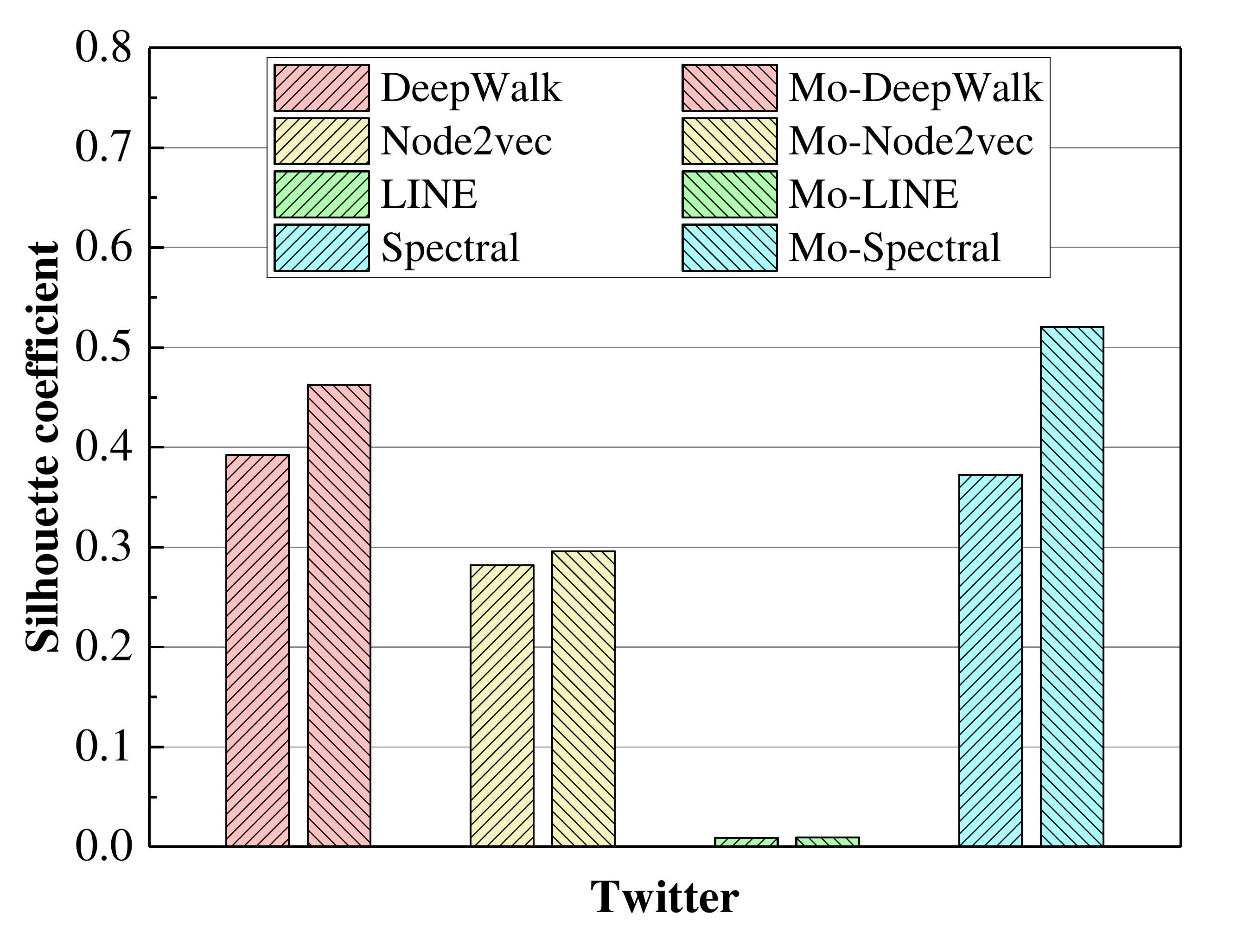}\label{fig:sctwitter}}
    \end{minipage}
    \begin{minipage}{0.23\textwidth}
        \centerline{\includegraphics[width=3.5cm]{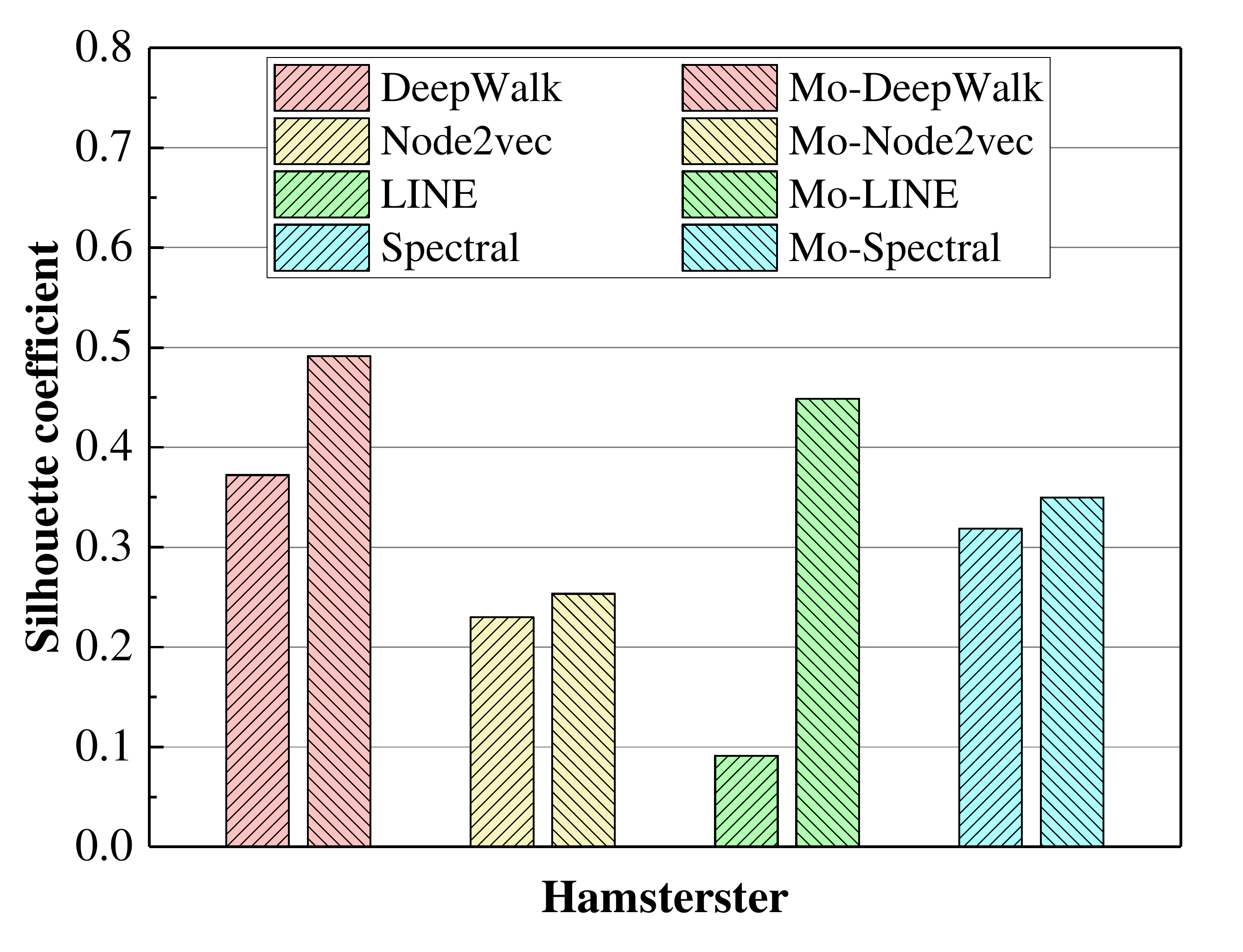}\label{fig:scham}}
    \end{minipage}
    \caption{Silhouette coefficient of the four networks.}
    \label{fig:sc}
\end{figure}
\textbf{Clustering. }
Herein, we use SC (Silhouette Coefficient) to evaluate the clustering results. SC is proposed with the combination of cohesion and separation.
A higher SC value reflects better clustering results. Specifically, SC equals avg$\{s(i)\}$, wherein, $s(i)$ is the SC of sample $i$ and it can be calculated by $s(i)=[b(i)-a(i)] /\max \{ a(i), b(i) \}$. Among them, $a(i)$ refers to the average distance of sample $i$ to other samples, which is considered as the evaluation metric for cohesion degree within the cluster. Correspondingly, $b(i)$ refers to the average distance of sample $i$ and all the nodes in another cluster that $i$ does not belong to. It is widely used to evaluate the separation degree between clusters.

\subsection{Results and Discussions}
\label{sec:rd}

Figure~\ref{fig:PAR} shows the experimental results of link prediction in the four datasets. We used DeepWalk~\cite{perozzi2014deepwalk}, Node2vec~\cite{grover2016node2vec}, LINE~\cite{tang2015line}, Spectral~\cite{zhang2019prone}, which are popular algorithms in research and engineering to optimize. The corresponding optimization algorithm is named Mo-DeepWalk, Mo-Node2vec, Mo-LINE, Mo-Spectral.

Figure~\ref{fig:PAR} (a)-(d) shows experimental results of four first-level indicators( \textbf{Precision}, \textbf{Recall}, \textbf{Accuracy} and \textbf{Specificity}). It has been observed that enhanced algorithms consistently yield superior performance. Meanwhile, the performance of these four metrics is close to each other.
For example, in Wiki, Mo-DeepWalk outperforms DeepWalk by about 4\% for all four metrics.  As for three other enhanced algorithms, more than 3\% performance gain is found.

Figure~\ref{fig:PAR} (e)-(f) shows experimental results of two second-level indicators: \textbf{F1-score} and \textbf{AUC}. \textbf{F1-score} can explore the balance of \textbf{Precision} and \textbf{Recall}. \textbf{AUC} avoids converting prediction probabilities into categories, and is not sensitive to whether the sample categories are balanced.
The figure shows that the proposed motif dimensional network representation learning has consistently improved all of the original algorithms. Such finding also indicates that the motifs represented the multivariate relationship can improve network representation. The improvement is significant on Wiki and Twitter.
The observation demonstrates that OFFER effectively improves the performance of the link prediction. 

Figure~\ref{fig:sc} shows the SC value comparison of the four enhanced algorithms against the original algorithms. It can be seen that in the four networks, the enhanced algorithms outperform with higher SCs. The mechanism of OFFER is that it can capture multivariate relationships. Among them, the clustering results of DeepWalk, Node2vec, and Spectral generally perform better than LINE does. This may be due to 2 labels used in the experiments, whereas LINE is usually applied for scenarios with more labels.

\section{Conclusion and Future work}
\label{sec:con}

In this work, we propose a framework called OFFER to enhance the performance of network representation algorithms.
We take four network representation algorithms as examples for optimization, including DeepWalk, Node2vec, LINE, and Spectral Clustering. And link prediction is employed to verify the performance of the enhanced network representation. Experimental results show that all enhanced algorithms outperform the original ones. Limitation of the OFFER framework includes its only focus on the network structure characteristics and its refinement of the learning process. Besides, only the triangle motif has been applied to represent multivariate relationships in the experiments, while higher-order motifs can capture relationships among more entities. But the price is a more complicated calculation and choosing a more suitable motif.

Multivariate relationships with higher-order embedding mechanism proposed in this work can be applied to graph topology inference is significant and interesting. Moreover, identifying a matching motif for a certain network also worth further investigation. Therefore, enhancing network representation learning of multivariate relationships is foreseeable future work.

\section{Acknowledgments}

The authors would like to thank Kaiyuan Zhang and Wenya Li for their support and help in the experiment.

\bibliographystyle{ACM-Reference-Format}
\bibliography{References}

%
%
%
%
%
%
%
%

\end{document}